\documentclass[10pt]{iopart}
\usepackage[colorlinks=true,citecolor=blue,urlcolor=blue]{hyperref}
\usepackage{amssymb}
\usepackage{bbm}
\usepackage{graphicx}
\usepackage{ulem}

\newcommand{\be}{\begin{equation}}
\newcommand{\ee}{\end{equation}}

\newcommand{\rmS}{{\rm S}}
\newcommand{\rmL}{{\rm L}}
\newcommand{\rmR}{{\rm R}}
\newcommand{\rmG}{{\rm G}}

\newcommand{\kBT}{k_{\rm B}T}
\begin{document}

\title[Single-electron thermal devices coupled to a mesoscopic gate]{Single-electron thermal devices coupled to a mesoscopic gate}

\author{Rafael S\'anchez}
\address{Instituto Gregorio Mill\'an, Universidad Carlos III de Madrid, 28911 Legan\'es, Madrid, Spain}
\author{Holger Thierschmann}
\address{Kavli Institute of Nanoscience, Faculty of Applied Sciences, Delft University of Technology, Lorentzweg 1, 2628 CJ Delft, The Netherlands}
\author{Laurens W. Molenkamp}
\address{Experimentelle Physik 3, Physikalisches Institut, Universit\"at W\"urzburg, Am Hubland, 97074 W\"urzburg, Germany}
%\ead{submissions@iop.org}
\vspace{10pt}
%\begin{indented}
%\item[]February 2014
%\end{indented}

\begin{abstract}
We theoretically investigate the propagation of heat currents in a three-terminal quantum dot engine.
Electron-electron interactions introduce state-dependent processes which can be resolved by energy-dependent tunneling rates. We identify the relevant transitions which define the operation of the system as a thermal transistor or a thermal diode. In the former case, thermal-induced charge fluctuations in the gate dot modify the thermal currents in the conductor with suppressed heat injection, resulting in huge amplification factors and the possible gating with arbitrarily low energy cost. In the latter case, enhanced correlations of the state-selective tunneling transitions redistribute heat flows giving high rectification coefficients and the unexpected cooling of one conductor terminal by heating the other one.
We propose quantum dot arrays as a possible way to achieve the extreme tunneling asymmetries required for the different operations.
\end{abstract}

%\maketitle

\section{Introduction}

The control of heat flows in electronic conductors is one of the present day technological challenges. Besides the conversion of heat currents into electrical power which is the main focus of thermoelectrics, there is the possibility of making all-thermal circuits that work only with heat currents and temperature gradients. For this purpose, all-thermal versions of electronic devices such as diodes and transistors need to be operative. 

Mesoscopic conductors are good candidates~\cite{giazotto_opportunities_2006,benenti_fundamental_2016}: their energy spectrum can be easily designed by using low dimensional structures showing quantum confinment (quantum dots, quantum wires\dots). Also, the presence of different gaps due to interaction effects (e.g. Coulomb blockade) or superconductor interfaces can be controlled. This possibility has been evident in the last years with the proposal and experimental realization of heat engines~\cite{staring_coulomb_1993,dzurak_observation_1993,godijn_thermopower_1999,
scheibner_thermopower_2005,scheibner_sequential_2007,svensson_lineshape_2012,
svensson_nonlinear_2013,thierschmann_diffusion_2013,hotspots,holger,sothmann_rectification_2012,roche_harvesting_2015,
review,sierra_strongly_2014,svilans_nonlinear_2016} and electronic refrigerators~\cite{molenkamp_peltier_1992,giazotto_opportunities_2006,
edwards_cryogenic_1995,prance_electronic_2009,nahum_electronic_1994,leivo_efficient_1996,
timofeev_electronic_2009,anna} based on nanoscale systems, and with the detection of heat currents~\cite{molenkamp_peltier_1992,chiatti_quantum_2006,meschke_single_2009,
jezouin_quantum_2013,riha_mode_2015,cui_quantized_2017}. Thermal rectifiers and transistors have also been proposed~\cite{wang_thermal_2007,segal_single_2008,segal_nonlinear_2008,
ruokola_thermal_2009,ming_ballistic_2010,ruokola_single_2011,giazotto_thermal_2013,
diode,sierra_nonlinear_2015,jiang_phonon_2015,vannucci_interference_2015,
benenti_from_2016,joulain_quantum_2016,guillem} and 
experimentally implemented~\cite{saira_heat_2007,scheibner_quantum_2008,martinez_efficient_2013,martinez_rectification_2015}. 

A crucial aspect for an electronic thermal device is how it interacts with its environment. Inelastic transitions in the device can be due to the coupling to fluctuations in the electromagnetic environment~\cite{ruokola_theory_2012,henriet_electrical_2015,guillem}, or to phononic~\cite{entin_three_2010,entin_enhanced_2015} or magnetic~\cite{sothmann_magnon_2012} baths. Remarkably in mesoscopic conductors, the dominant interaction can be engineered by introducing additional components which mediate the coupling to the environment, e.g. a quantum point contact~\cite{khrapai_double-dot_2006}, quantum dots~\cite{hotspots,holger,hartmann_voltage_2015}, or photonic cavities~\cite{bergenfeldt_hybrid_2014,partanen_quantum_2016,hofer_quantum_2016}. This allows one to define different interfaces for the different operations. From a theoretical point of view, this permits the environment to be described as a third terminal, and the auxiliary system (mediating the system-environment coupling) to be treated in equal footing as the conductor.

\begin{figure}[b]
\begin{center}
\includegraphics[width=\linewidth,clip]{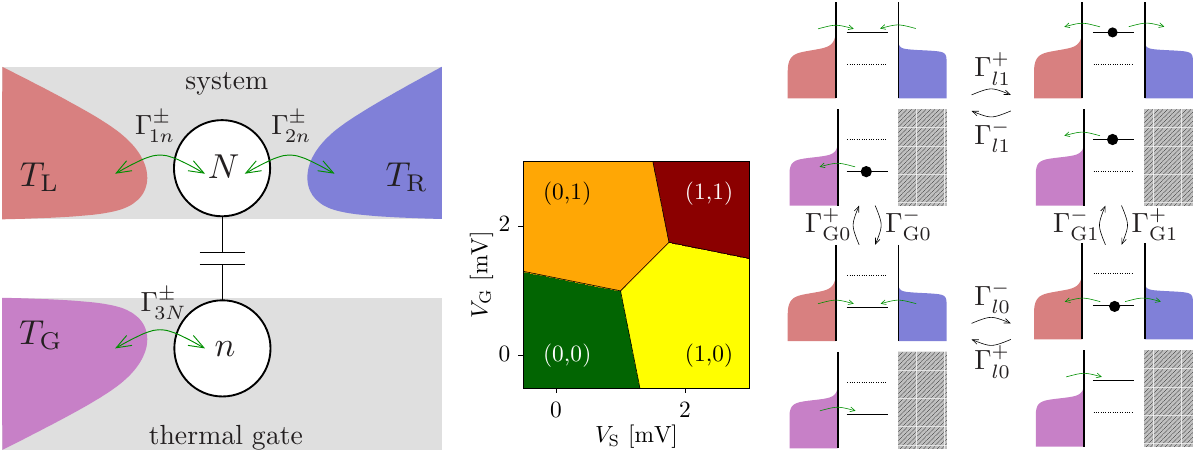}
\end{center}
\caption{\label{occupation}A quantum dot thermal gate. The heat flows in a two terminal electronic conductor are modulated by the temperature $T_\rmG$ of a third reservoir. The coupling is mediated by the Coulomb interaction at two quantum dots. Energy-dependent tunneling rates are affected by the charging of the other dot. (Center): Charge stability diagram of the system as the quantum dot gate voltages are tuned in the region $(N,n)$ with $N,n$=0,1. We have considered $\alpha_i=0.1$ and $\beta=0.02$. (Right): Tunneling processes involved in the dynamics of the system. The rates $\Gamma_{im}^\pm$ for tunneling in ($+$) or out ($-$) of each quantum dot depend on the occupation of the other dot, $m$.
}
\end{figure}

Here we propose a multi-terminal system of two capacitively coupled quantum dots as a versatile configuration for the efficient manipulation of electronic heat currents, as sketched in Fig.~\ref{occupation}. It works as a bipartite system: One of them is coupled to two terminals and is considered as the conductor whose heat currents are to be manipulated. The other dot is tunnel-coupled to a third terminal and serves as a gate system. Energy exchange between the two systems is mediated by electron-electron interactions~\cite{heij_negative_1999,michalek_current_2002,chan_strongly_2002,
hubel_two_2007,goorden_two_2007}.
Effects such as mesoscopic Coulomb drag~\cite{drag,bischoff_measurement_2015,kaasbjerg_correlated_2016,
keller_cotunneling_2016}, enhanced correlations~\cite{mcclure_tunable_2007,goorden_cross_2008,michalek_dynamical_2009,
correlations}, heat engines~\cite{hotspots,holger,sothmann_rectification_2012,roche_harvesting_2015,detection,
donsa_double_2014,zhang_three_2015,dare_powerful_2017}, rectification of noise~\cite{hartmann_voltage_2015,pfeffer_logical_2015}, non-local power generation~\cite{whitney_thermoelectricity_2016}, relaxation time scales~\cite{schulenborg_detection_2014}, or the relation of information and thermodynamics~\cite{strasberg_thermodynamics_2013,horowitz_thermodynamics_2014,koski_on-chip_2015} have been investigated in similar configurations. Thermoelectric effects in three terminal configurations have been described where the coupling of a junction to the heat source is mediated by other kinds of interactions~\cite{khrapai_double-dot_2006,entin_three_2010,ruokola_theory_2012,sothmann_magnon_2012,
review,entin_enhanced_2015,
henriet_electrical_2015,hofer_quantum_2016}. 

Energy exchange mediated by electrostatic coupling has the additional advantage of allowing for defining thermal insulating system-gate interfaces. This way eventual heat currents leaking from the gate to the conductor, e.g. due to phonons, are suppressed. Also, phononic heat currents along the conductor are not affected by charge fluctuations in the gate and hence are not expected to affect the thermal gating effects. Furthermore, in the low temperature regime discussed here, the contribution of phonons in the conductor can be neglected.

In a recent experiment~\cite{thgating}, the gating of voltage-induced electronic currents in the conductor by the modulation of the temperature of the gate was observed. We extend the investigation of such thermal gating to the gating of thermally-induced heat currents. A first step was done in Ref.~\cite{short} where the gate dot acts as a switch. As we show here, exploiting the energy dependence of tunneling effects in the conductor dot, the same device becomes a versatile thermal device. We connect the different operations of the device to different asymmetries. Energy-filtered couplings lead to a thermal transistor (Sec.~\ref{sec:transistor}). Combined energy-dependent and left-right asymmetries give a thermal diode as well as the effect of cooling by heating (Sec.~\ref{sec:diode}). We discuss possible implementations of all these asymmetries and how to tune them with quantum dot arrays in Sec.~\ref{sec:filters}.

In the remaining of the text, Sec.~\ref{sec:currents} describes the theoretical model and the heat currents and Sec.~\ref{sec:gating} introduces the thermal gating effect. Conclusions are discussed in Sec.~\ref{sec:conclusion}.

%\begin{figure}[b]
%\begin{center}
%\includegraphics[width=0.9\linewidth,clip] {./figs/dqdscheme.eps}
%\end{center}
%\caption{\label{scheme} Sketch of a quantum dot thermal gate. The heat flows in a two terminal electronic conductor are modulated by the temperature $T_\rmG$ of a third reservoir. The coupling is mediated by the Coulomb interaction at two quantum dots. Energy-dependent tunneling rates are affected by the charging of the other dot.
%}
%\end{figure}

\section{Heat currents}
\label{sec:currents}

Our model is based on the interaction of two mesoscopic systems: the conductor system and the thermal gate. For capacitively coupled quantum dots, the interaction is given in terms of the electrostatic repulsion of electrons occupying them. For simplicity, we will assume that it is very large for electrons in the same dot, so we restrict our considerations to single-electron occupations in each quantum dot. Interdot Coulomb interaction is described by a constant $E_C$ given by the geometric capacitances of the system.~\cite{hotspots} 

Through all this work, we consider non-equilibrium situations only due to temperature gradients $\Delta T_i=T_i-T$ applied to one terminal with respect to the reference temperature, $T$. We do not restrict to the linear regime where gradients are small. No bias voltage is applied. Index $l=\rmL,\rm R$ will always refer to the conductor terminals, while $i$ labels any terminal. G will denote the gate. The internal potential $U_{im}$ of each quantum dot $i$ is then shifted by $E_C$ when the charge of the other dot $m$ fluctuates: $U_{i1}=U_{i0}+E_C$. Hence every energy exchange between the system and the gate is given in terms of this quantity.

In an experimental realization, the internal potentials are tuned by means of gate voltages which we include with their corresponding lever arms $\alpha_i$ and $\beta$: $U_{{\rm S}0}=\varepsilon_\rmS+\alpha_\rmS eV_\rmS+\beta eV_\rmG$, and $U_{\rmG 0}=\varepsilon_\rmG+\alpha_\rmS eV_\rmG+\beta eV_\rmS$. This way, the charge configuration of the system can be externally manipulated by moving $U_{im}$ with respect to the Fermi energy of the leads, $\mu_i$. This is shown in Fig.~\ref{occupation}: each region of this stability diagram is dominated by one of the occupations $(N,n)$, with $N,n=0,1$, where $N$ represents the occupation of the system and $n$ denotes the occuppation of the  gate. Charge fluctuates around the intermediate regions. 

Two triple points appear where charge fluctuations are present in both dots. We call the region between them the stability vertex, whose size is given by $E_C$. Thermoelectric heat engines based on the correlation of the system and the gate operate in this region~\cite{hotspots,holger,koski_on-chip_2015}.

%\begin{figure}[t]
%\begin{center}
%\includegraphics[width=0.4\linewidth,clip] {cycle.eps}
%\end{center}
%\caption{\label{cycle}Tunneling processes involved in the dynamics of the system. The rates $\Gamma_{im}^\pm$ for tunneling in ($+$) or out ($-$) of each quantum dot depends on the occupation of the other dot, $m$.
%}
%\end{figure}

We write a rate equation for the occupation of the charge states $\lambda=(N,n)$:
\be
\dot P(\lambda)=\sum_{\lambda'}\left[\Gamma_{\lambda\leftarrow \lambda'}P(\lambda')-\Gamma_{\lambda'\leftarrow \lambda}P(\lambda)\right],
\ee
with the tunneling rates $\Gamma_{(1,n)\leftarrow(0,n)}=\sum_{l}\Gamma_{ln}^+$ (with $l=\rmL,\rm R$), and $\Gamma_{(N,1)\leftarrow(N,0)}=\Gamma_{\rmG N}^+$. For the reversed processes, we replace $+\rightarrow-$. They are written as $\Gamma_{im}^\pm=\Gamma_{im}f_i^\pm(\Delta U_{im})$, where the transparencies of the barrier $\Gamma_{im}$ depend on the occupation $m$ of the other dot, $f_i^+(E)=[1+\rme^{E/\kBT_i}]^{-1}$ is the Fermi function, and $f_i^-(E)=1-f_i^-(E)$. We have defined $\Delta U_{im}=U_{im}-\mu_i$. We assume weak dot-lead couplings, $\Gamma_{im}\ll\kBT$, where transport is dominated by sequential tunneling events. A schematic representation of the possible trajectories is represented in Fig.~\ref{occupation}.

The dc heat currents are obtained by the steady state occupations $\bar P(\lambda)$ satisfying: $\dot{\bar P}(\lambda)=0$. They are given by~\cite{hotspots}:
\begin{eqnarray}
J_l&=\sum_n\Delta U_{ln}\left[\Gamma_{ln}^+\bar P(0,n)-\Gamma_{ln}^-\bar P(1,n)\right]\label{eq:jl}\\
J_\rmG&=\sum_N\Delta U_{\rmG N}\left[\Gamma_{\rmG N}^+\bar P(N,0)-\Gamma_{\rmG N}^-\bar P(N,1)\right].\label{eq:jG}
\end{eqnarray}
They are defined as positive when heat flows out the terminal. From the previous expressions, one can straightforwardly write the state-resolved currents $J_{ln}$ and $J_{\rmG N}$ such that: $J_{i}=\sum_mJ_{im}$~\cite{detection}. As we are applying no bias voltage, heat is conserved, and $\sum_iJ_i=0$. In the following, we write $J_i(\Delta T_j)$ as the heat current injected from terminal $i$ in response to a temperature gradient applied to terminal $j$, with all other terminals being (except when explicitly stated) at temperature $T$.

\begin{figure}[t]
\begin{center}
\includegraphics[width=0.7\linewidth,clip]{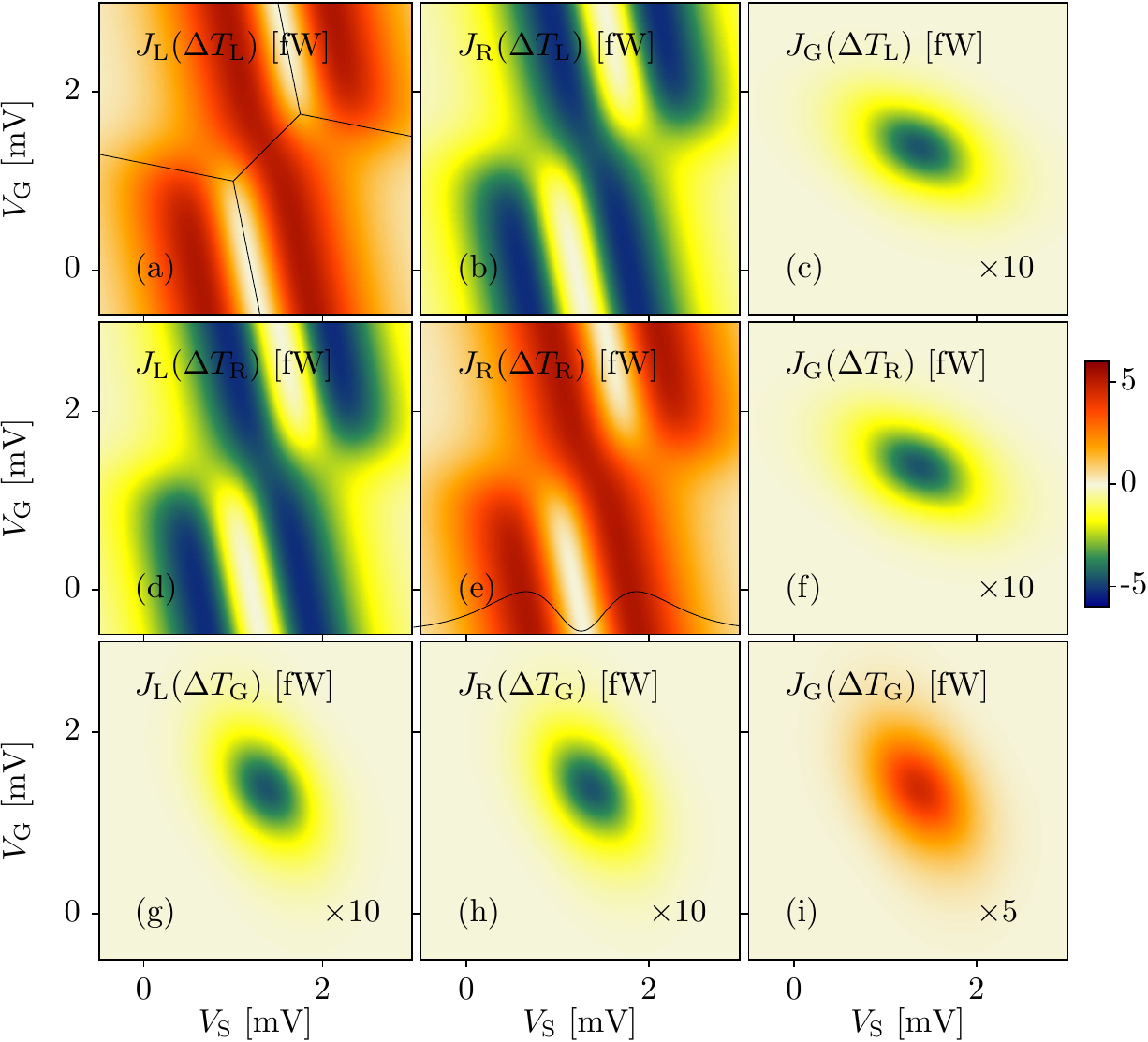}
\end{center}
\caption{\label{currentsSym}Heat currents in response to the different temperature gradients as functions of the quantum dot gate voltages. In panel (a) we mark the limits of the charge stability regions. Heat current in the gate dot is only finite close to the two triple points. As a guide to the eye, panel (e) includes a cut of $J_\rmR(\Delta T_\rmR)$ at $V_\rmG=-0.5$~mV showing the double peak structure. Parameters: $E_C=90$~meV, $\kBT=0.243$~K, $\Delta T_i=\kBT/2$, $\Gamma_{lN}=10$~$\mu$eV, $\Gamma_{\rmG n}=1$~$\mu$eV.
}
\end{figure}

The discrete level of each quantum dot filters the energy of the tunneling processes. The flow of heat is restricted to the region where they are close to the Fermi energy, as shown in Fig.~\ref{currentsSym}. When the temperature gradient is applied longitudinally in the conductor (i.e. either in the left or right terminal), the conductor heat currents present a double peak around the crossing of the level with  the Fermi energy. At that particular point far from the stability vertex (such that fluctuations in the gate are frozen), $J_l=0$ because of particle-hole symmetry. This is a well-known feature of two-terminal quantum dots~\cite{vavilov_failure_20015}. In our system, such a signal is broken at the stability vertex as a consequence of the charging of the gate dot.

The mechanism for heat transport in the gate is different. It relies on charge fluctuations in the two dots which involve the four charging states in a closed loop trajectory~\cite{hotspots}. In a sequence of the form $(0,0){\leftrightarrow}(1,0){\leftrightarrow}(1,1){\leftrightarrow}(0,1){\leftrightarrow}(0,0)$, the events of tunneling in and out of the gate dot occur at different energies due to the different occupations of the conductor dot. The energy difference, $\pm E_C$, is transferred between the two-terminal conductor and the gate. The sign of the transfer depends on the direction of the above sequence. Hence $J_\rmG(\Delta T_j)$ shows up as a spot in the middle of the stability vertex~\cite{koski_on-chip_2015}, as shown in Fig.~\ref{currentsSym}. 

In agreement with the Clausius statement of the second law, heat flows out of the hot reservoirs, as shown in the diagonal panels starting from the top left in Fig.~\ref{currentsSym}. We note that in the symmetric configuration shown there, the absorption of this heat current is shared by the corresponding other (cold) terminals, where currents are negative. As we discuss below, this is not necessarily the case in the presence of energy-dependent rates.

\section{Thermal gating}
\label{sec:gating}

The effect of the gate dot on the heat flow in the conductor is twofold: it exchanges heat with the conductor and affects the heat flow {\it within} the conductor. The former one modifies energy conservation. The second is due to the shift in the position of the energy level of the conductor when  the occupation of the gate dot changes. This breaks the double peak structure shown in Fig.~\ref{currentsSym}. The effect is evident in the stability vertex where transport fluctuations in the conductor coincide with charge fluctuations in the gate dot~\cite{correlations}. These occur at a higher rate when the gate reservoir is hot. Hence tuning the temperature $T_\rmG$ modifies the transport properties of the conductor. We call this the thermal gating effect. 

In the electric response, it manifests as a clover-leaf structure with positive and negative regions around the stability vertex~\cite{thgating,comptes}. The sign can be understood in terms of dynamical channel  blockade: It depends on whether the fluctuations in the gate dot contributes to open or close the relevant transport channels~\cite{heij_negative_1999,michalek_current_2002,mcclure_tunable_2007,stacks,michalek_dynamical_2009}.

\begin{figure}[t]
\begin{center}
\includegraphics[width=0.7\linewidth,clip]{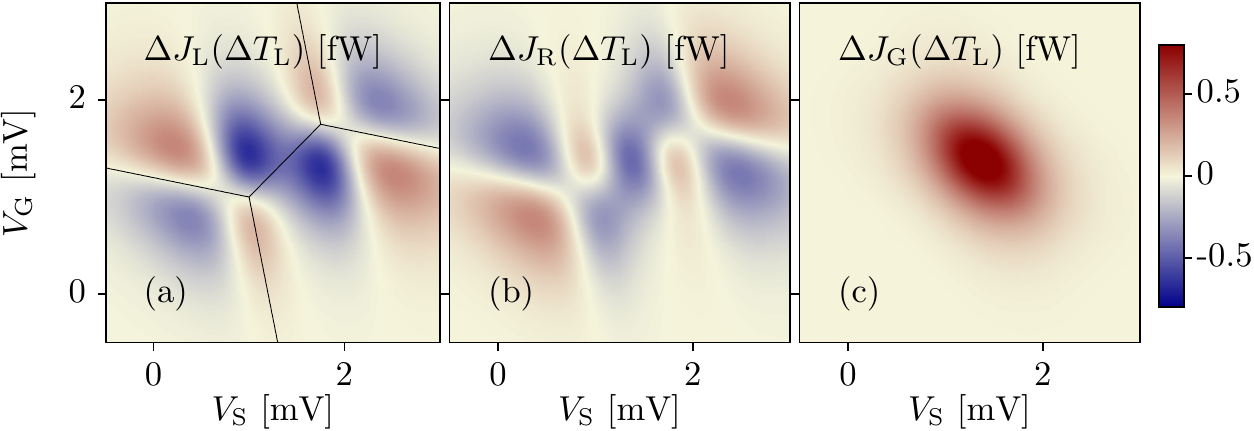}
\end{center}
\caption{\label{thgateSym}Thermal gating of heat currents when increasing the gate temperature $\Delta T_\rmG=\kBT/2$ for the same symmetric configuration as in Fig.~\ref{currentsSym}. Note that $\Delta J_l(\Delta_m)=0$ along the conditions where the fluctuations between ($N$,0) and ($N$,1) are balanced.
}
\end{figure}

Here we are interested in the thermal gating of heat currents. We quantify it by defining the modulation:
\be
\Delta J_i(\Delta T_j)=J_i(\Delta T_j,\Delta T_\rmG)-J_i(\Delta T_j,0).
\ee
We plot it in Fig.~\ref{thgateSym} for a symmetrical configuration with $\Gamma_{ln}=\Gamma$. For heat currents, we observe a clover-leaf structure centered at each triple point. Interestingly, $\Delta J_l=0$ along the conditions $P(N,0)=P(N,1)$. There, the level of the gate dot is aligned with the Fermi energy such that charging/uncharging the gate dot is independent of the lead temperature. 

The clover-leaf structures are the effect of thermal gating due to fluctuations in the gate. However, in the center of the stability vertex we also observe signatures of the additional effect of energy exchange with the gate system. Obviously, if we inject heat from a third terminal, the total heat flow in the conductor is affected. The performance of our system as an all-thermal device, e.g. a thermal transistor, requires the suppression of such contributions. For example, reducing the transparency $\Gamma_{\rmG N}$ reduces $J_\rmG$, leaving the clover leaf structure almost unaffected~\cite{comptes}. This is an indication that based on fluctuations of the gate only one can have a thermal gating effect, in principle involving no heat injection into the conductor.

\section{Energy filtering: A thermal transistor}
\label{sec:transistor}

An all-thermal transistor modifies the heat current in a conductor due to changes in the temperature of a third (gate) terminal. This is the thermal gating effect discussed in the previous section. In this case, the two conductor terminals act as the emitter and the collector, while the gate terminal is considered as the base. As for an electric transistor, we can define the thermal amplification factor as~\cite{jiang_phonon_2015,joulain_quantum_2016}:
\be
\alpha_{lm}=\frac{|\Delta J_l(\Delta T_m)|}{|\Delta J_\rmG(\Delta T_m)|},
\ee
where $\Delta J_\rmG(\Delta T_m)$ is the change of heat injected from the gate when tuning $T_\rmG$. For the system to work as a thermal transistor, one needs $\alpha_{lm}\gg1$

The challenge is to achieve a measurable thermal gating by injecting a small amount of heat from the gate. In our system, heat is injected only via electron-electron interactions and we can therefore control it by selecting the charge of the system. As discussed above, the mechanism for heat transfer in our system is based on the occurance of charge fluctuations involving all the four charging states in a loop, as presented in Fig.~\ref{occupation}. In this section we discuss how to avoid such trajectories. We will see that this is the most effective way to suppress the energy exchange between the system and the gate, while still having a gating effect. This can be achieved either by operating at low tempreature, $kT\ll E_C$ (i.e. eliminating one of the states), or by acting on the energy-resolved tunneling rates (eliminating one of the transitions)~\cite{short}. Time-resolved measurement of fluctuations in the former configuration have been recently reported in single-electron transistors~\cite{singh_distribution_2016}. 

We focus here on the second case, where the tunneling rates in the conductor are strongly energy dependent, for instance $\Gamma_{l0}\ll\Gamma_{l1}$. This allows us to consider the regime $\kBT\sim E_C$. An energy  dependence in tunneling rates is expected to occur naturally as in quantum mechanics the tunneling probability depends exponentially on the energy of the electron with respect to the height of the barrier. Furthermore, in semiconductor quantum dots, the energy dependence of the barrier transparencies can be modulated experimentally by means of gate voltages~\cite{holger}. Additionally, a more complicated quantum dot composition can be used, where only two dots provide the system-gate interface and the other ones are used as energy filters. We discuss several such geometries in Sec.~\ref{sec:filters}.

In the limit $\Gamma_{l0}=0$, the transitions (0,0)$\leftrightarrow$(1,0) are avoided. Hence the state $(1,0)$ is only populated through fluctuations of the form (1,1)$\leftrightarrow$(1,0) taking place in the gate dot. Currents in the conductor are conditioned on the occupation of the gate dot. When it is empty, the transport transitions are blocked. We can then write: $J_l=J_{l1}$ and $J_{l0}=0$. The gate dot becomes a switch. 

In this case, there are no transitions that correlate fluctuations at the two dots. Then, state-dependent currents are conserved in each conductor: $J_{l0}=0$, $J_{\rmL1}+J_{\rmR1}=0$, and $J_{\rmG N}=0$~\cite{detection}. From the latest and Eq.~(\ref{eq:jG}), we get the relations:
\be
\bar P(N,0)=\rme^{\Delta U_{\rmG N}/\kBT_\rmG}\bar P(N,1),
\ee
emphasizing that the gate satisfies detailed balance. Using the level-resolved conservation laws given above, this model admits a simple analytical solution without needing to solve the master equation. The heat currents in the conductor are:
\be
\label{jltrans}
J_l=\Delta U_{l1}\Gamma_{{\rm seq}1}(f_{l1}-f_{\bar l1})\langle n(T_\rmG)\rangle,
\ee
where we have written $\Gamma_{{\rm seq}1}^{-1}=\sum_l\Gamma_{l1}^{-1}$, $f_{l1}=f_l(\Delta U_{l1})$, and $\bar l$ is the opposite terminal to $l$ in the conductor. The effect of the gate enters only in the average occupation $\langle n(T_\rmG)\rangle=P(0,1)+P(1,1)$ given by:
\be
\langle n(T_\rmG)\rangle=\frac{\sum_l\Gamma_{l1}f_{\rmG0}f_{\rmG1}}{\sum_l\left(\Gamma_{l1}^+f_{\rmG0}+\Gamma_{l1}^-f_{\rmG1}\right)}. 
\ee
It modifies the current expected for an isolated single channel at energy $\varepsilon$:
\be
J_{{\rm sc},l}(\varepsilon)=\varepsilon\Gamma_{\rm seq}(\varepsilon)[f_{l}(\varepsilon-\mu)-f_{\bar l}(\varepsilon-\mu)].
\ee

\begin{figure}[t]
\begin{center}
\includegraphics[width=0.7\linewidth,clip] {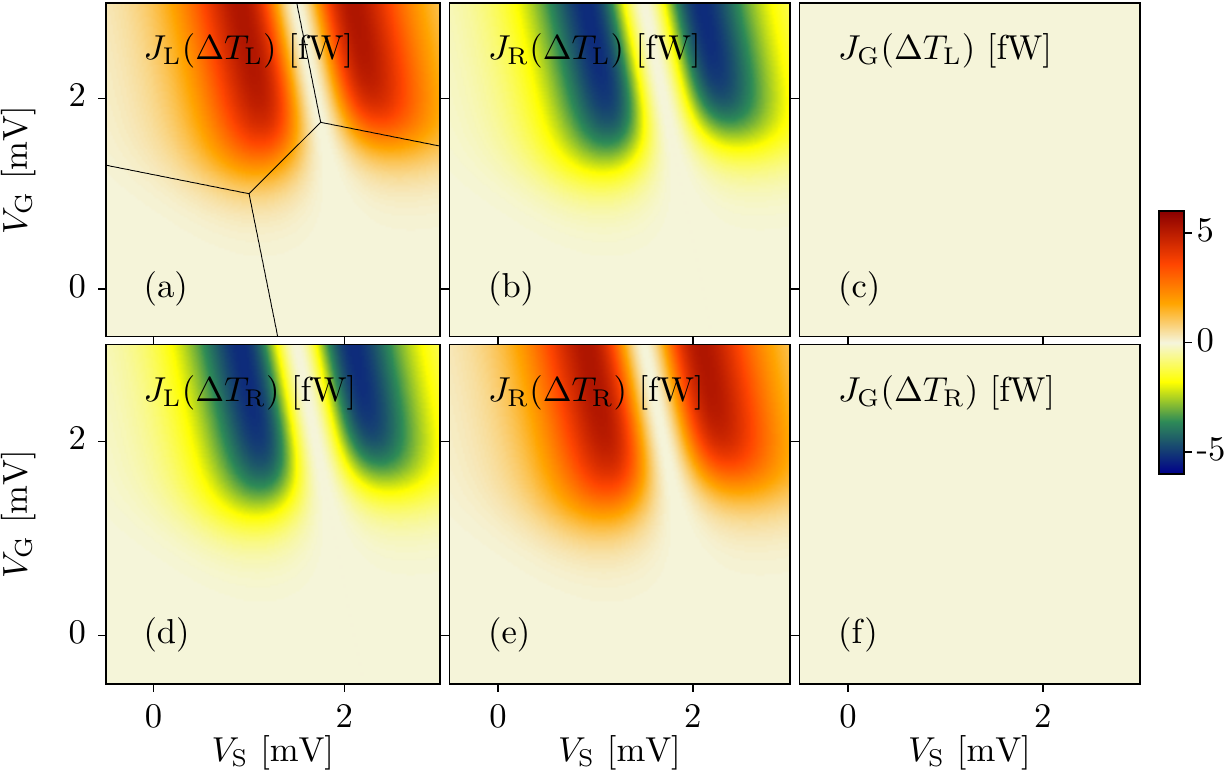}
\end{center}
\caption{\label{currentsselU}Heat currents in response to the different temperature gradients as functions of the quantum dot gate voltages in the energy-selected-tunneling configuration, with $\Gamma_{\rmL0}=\Gamma_{\rmR0}=0$. Currents are hence confined to the region with $n_\rmG=1$. Note that there is no heat injected from the gate dot: $J_\rmG=0$. Also, the gradient in the gate terminal does not lead to any heat current: $J_l(\Delta T_\rmG)=0$, so we do not plot them here. The rest of parameters is as in Fig.~\ref{currentsSym}.
}
\end{figure}

This effect can be clearly observed in Fig.~\ref{currentsselU}, where all possible currents in response to a temperature gradient applied to the different terminals of the conductor are plotted. By comparing it with the symmetric (energy independendent) rates configuration shown in Fig.~\ref{currentsSym}, we observe that transport is suppressed in the lower region of the stability diagram, where $\langle n\rangle\approx0$ as a result of the asymmetric tunneling rates. 

\begin{figure}[t]
\begin{center}
\includegraphics[width=0.7\linewidth,clip] {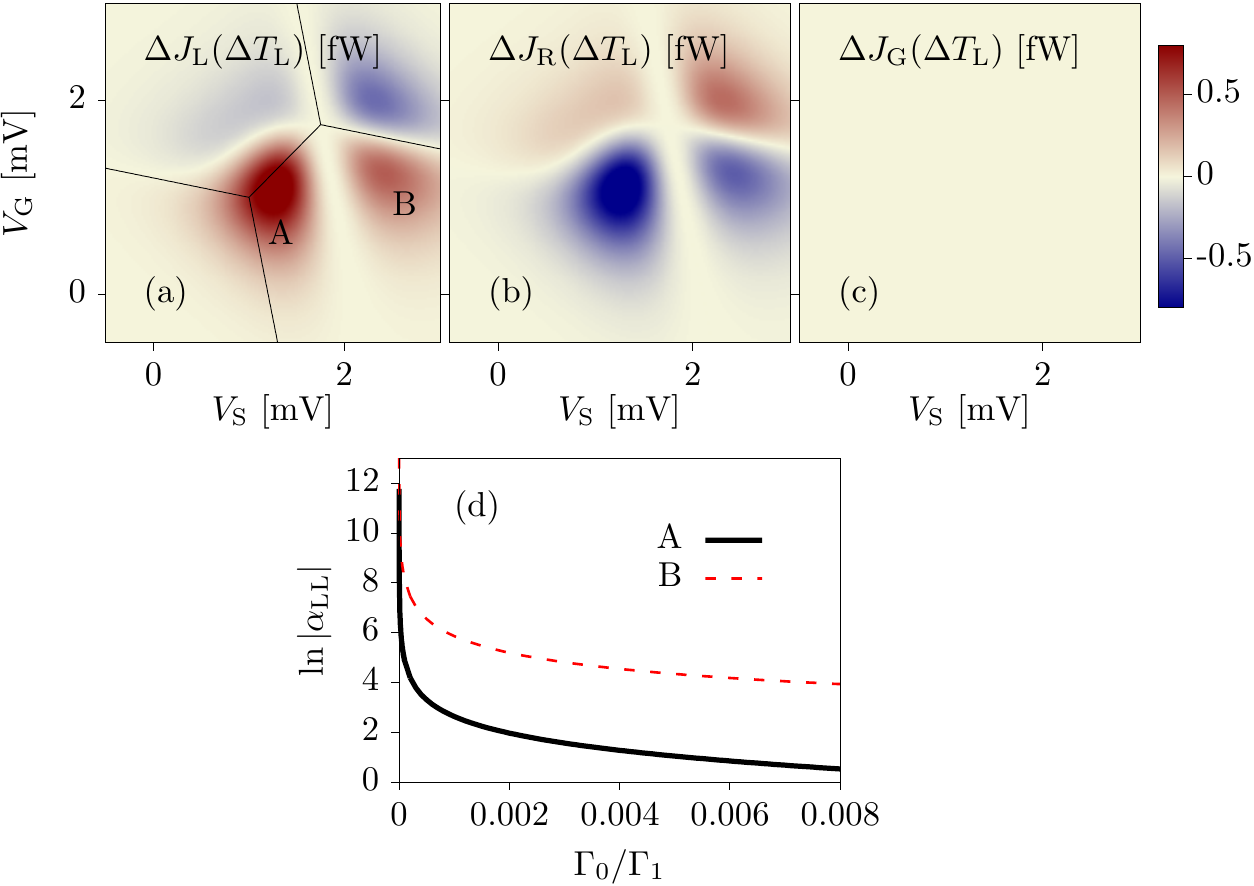}
\end{center}
\caption{\label{thgateSelU} (a-c) Thermal gating of heat currents when increasing the gate temperature $\Delta T_\rmG=\kBT/2$ for the same configuration as in Fig.~\ref{currentsselU}. (d) Thermal amplification coefficient as a function of the tunneling rate state-dependence. The two curves correspond to the maximal thermal gating signal $\Delta I_l^\rmL(\Delta T_\rmL)$ in (a). As the rates $\Gamma_{l0}=\Gamma_0$ (corresponding to an empty gate  dot) vanish, the amplification increases and the systems behaves as an ideal thermal transistor.
}
\end{figure}

We note that no heat current flows from the gate into the conductor. Nevertheless, the thermal gating effect exists (cf. Fig.~\ref{thgateSelU}) due to the dynamical blockade of the heat currents: 
\be
\Delta J_l(\Delta T_m)=J_{{\rm sc},l}(\Delta U_{l1})[\langle n(T_\rmG+\Delta T_\rmG)\rangle-\langle n(T_\rmG)\rangle].
\ee
In this case only the clover-leaf around the triple point close to (1,1) appears. Also, $J_\rmG=0$ for any configuration, as expected, and $\Delta J_\rmG(\Delta T_\rmL)=0$. As a consequence, the thermal amplification factor diverges (within our approximations) and the system operates as an ideal transistor. Note that the leading order in an expansion of the thermal gating for this case is quadratic in the temperature gradient: $\Delta J_l(\Delta T_j)\sim\Delta T_j\Delta T_\rmG$.

%Differently to other proposals where heat injection from the base cannot be controlled~\cite{jiang_phonon_2015}, this effect does not rely on non linear expansions.

%\begin{figure}[t]
%\begin{center}
%\includegraphics[width=0.7\linewidth,clip] {./figs/amplifLrG_Gg1DT1.eps}
%\end{center}
%\caption{\label{amplifL}Thermal amplification coefficient as a function of the tunneling rate state-dependence. The two curves correspond to the maximal thermal gating signal $\Delta I_l^\rmL(\Delta T_\rmL)$ in Fig.~\ref{thgateSym}. As the rates $\Gamma_{l0}=\Gamma_0$ (corresponding to an empty gate  dot) vanish, the amplification diverges and the systems behaves as an ideal thermal transistor.
%}
%\end{figure}

\subsection{Leakage currents}
Of course, in a real setup the transition rates cannot be exactly suppressed. In that case, heat leaks from the gate and the amplification coefficient becomes finite. We take this into account in Fig.~\ref{thgateSelU}, where $\alpha_{\rmL\rmL}$ is calculated at the two maxima of $\Delta J_\rmL(\Delta T_\rmL)$ (labeled as A and B in Fig.~\ref{thgateSelU}(a)) for finite values of $\Gamma_{l0}=\Gamma_0$. For $\Gamma_0<\Gamma_1/100$, the amplification factor is several orders of magnitude larger than 1. Note that in the presence of a leakage current, the maximum at B (which is further from the stability vertex and is hence less affected by the leakage) gives a larger amplification factor, even if $\Delta J_\rmL(\Delta T_\rmL)|_A>\Delta J_\rmL(\Delta T_\rmL)|_B$ for $\Gamma_{0}=0$.

Being closer to the relevant triple point where sequential tunneling transitions are dominant, B is also expected to be less sensitive to higher-order tunneling processes (neglected here). These involve energy transfer between the two systems and hence avoid the divergence also in the ideal case $\Gamma_{l0}=0$. 

\subsection{Thermal gating without heating}
As we discuss above, a thermal transitor can work without any injection of heat in the conductor. This does not necessarily mean that it works without any energy cost: in order to increase $T_\rmG$ one in principle has to inject heat into the gate terminal. 

\begin{figure}[t]
\begin{center}
\includegraphics[width=0.7\linewidth,clip] {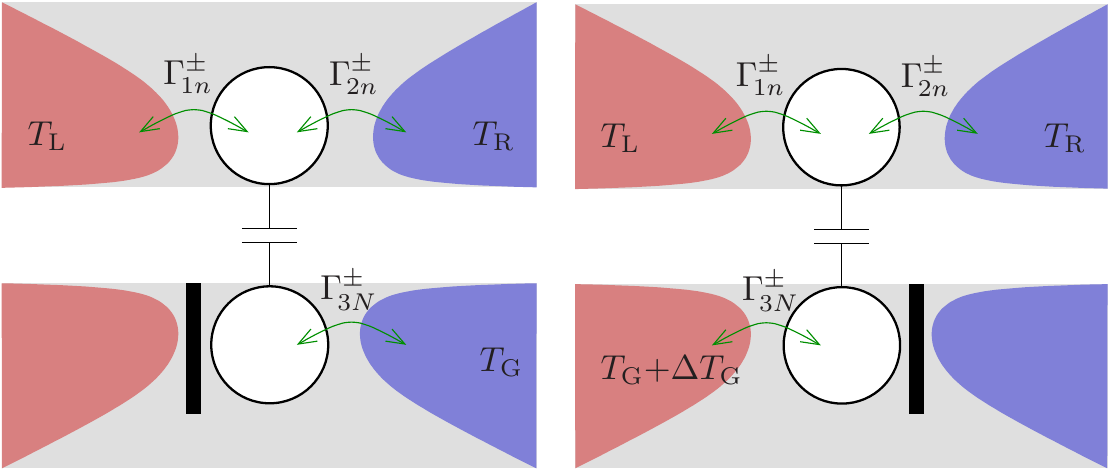}
\end{center}
\caption{\label{transistscheme}Gate thermal transistor. The temperature that  defines the gate fluctuations can be controlled by putting the gate quantum dot in contact with either a cold or a hot reservoir. This can be done by just opening or closing the corresponding tunneling barriers.
}
\end{figure}

In this respect, single-level quantum dots are also beneficial: The state of the dots is highly non-thermal and no temperature of the dot can be defined~\cite{whitney_thermoelectricity_2016}. The dynamics of the system depends on the non-equilibrium charge fluctuations in the quantum dots. The relevant gate temperature is that of the reservoir to which the quantum dot is coupled. We can then think of a configuration such that the gate dot is tunnel-coupled to two terminals, one cold and one hot, as depicted in Fig.~\ref{transistscheme}. By alternatively opening and closing one of the barriers, the charge fluctuations will adapt to the temperature of only one of them. Increasing $\Delta T_\rmG$ can then be done at an arbitrarily low energy cost (that of tuning the gates that control the barriers). 

Furthermore, the cold and hot reservoirs in the gate can even be in thermal contact with the conductor terminals. This way, only one thermal gradient is needed: the one that drives the heat currents in the conductor. The system is then reduced to two terminals connected in parallel by two interacting quantum dots: one supports a heat current. The other one is put in contact with only one of the reservoirs at a time. The current thus depends on to which side the gate dot is coupled.

\section{Combined state-dependent and mirror asymmetric tunneling: Thermal rectification}
\label{sec:diode}

The energy-dependent tunneling rates introduced by the coupling to the gate dot may also generate left-right asymmetries in the propagation of heat in the conductor: the  energy at which electrons will more probably tunnel through the two barriers will be different. Transport would then depend on where the temperature gradient is applied, leading to a thermal rectification effect, if $J_\rmL(\Delta T_\rmR)\neq J_\rmR(\Delta T_\rmL)$. %The  third terminal contributes to this effect also by serving as a heat sink.

A conductor that exploits this property by allowing the heat to flow for a forward, but not for a backward  gradient is a thermal diode. It is characterized by the thermal rectification coefficient, defined as:
\be
{\cal R}=\frac{\left|J_\rmL(\Delta T_\rmR)-J_\rmR(\Delta T_\rmL)\right|}{\left|J_\rmL(\Delta T_\rmR)\right|+\left|J_\rmR(\Delta T_\rmL)\right|}.
\ee
An ideal thermal diode operates at ${\cal R}\approx1$. We emphasize that, due to the presence of the gate terminal, heat is not conserved in the conductor subsystem and therefore ${\cal R}$ requires two currents to be defined. A symmetric configuration as the one considered in Fig.~\ref{currentsSym} obviously gives ${\cal R}=0$. 

Finite rectification coefficients are obtained for a simple left-right asymmetric configuration, $\Gamma_{ln}=\Gamma_l$. This case can be understood by considering for example the case $\Gamma_\rmL\gg\Gamma_\rmR$. Electrons in the left terminal will easily get into the dot, but once there, it takes a longer time to tunnel through the right barrier. During this time, it is in contact with the gate reservoir. On the other hand, left moving electrons hardly enter the dot, but then they can rapidly tunnel to the left. Therefore, the time that left- and right-moving electrons expend in contact with the gate reservoir is different, and so is the probablility that they lose heat in it. As a consequence, the heat current arriving to the left terminal when heating the right one will be larger. The role of the gate is totally pasive in this case, only acting as a heat sink. 

A configuration that emphasizes the left-right asymmetric energy-dependent rates suppresses tunneling processes through different barriers at different energies, for instance: $\Gamma_{\rmL0}=\Gamma_{\rmR1}=\Gamma$, and $\Gamma_{\rmL1}=\Gamma_{\rmR0}=x\Gamma$. Let us first consider for simplicity the case $x=0$. Then transitions $(0,n)\leftrightarrow(1,n)$ only occur through the left barrier for $n=0$ and through the right one, for $n=1$. In Sec.~\ref{sec:filters} we propose an implementation of this configuration. It also corresponds to an optimal (Carnot efficient) heat engine~\cite{hotspots}, and provides a maximal correlation between the different flows~\cite{correlations}: 
\begin{equation}
\label{currsdiode}
\frac{J_\rmL}{\Delta U_{{\rm S}0}}=-\frac{J_\rmR}{\Delta U_{{\rm S}1}}=\frac{J_\rmG}{E_C}={\cal I},
\end{equation}
where
\begin{equation}
\label{partdiode}
{\cal I}=\frac{\Gamma_{\rmL0}^+\Gamma_{\rmR1}^+\Gamma_{\rmG0}^+\Gamma_{\rmG1}^+}{\gamma^3}\rme^\frac{\Delta U_{\rmG0}}{\kBT_\rmG}\left({\rm e}^\frac{\Delta U_{{\rm s}1}}{\kBT_\rmR}{-}\rme^\frac{\Delta U_{{\rm s}0}}{\kBT_\rmL}e^\frac{E_C}{\kBT_\rmG}\right)
\end{equation}
is the particle current. The denominator $\gamma^3$ is determined by the normalization of the occupation probabilities. Note that if only one terminal is hot, ${\cal I}\neq0$ for finite $E_C$.
An electron can only be transported accross the conductor if it exchanges an energy $E_C$ with the gate. This property, which is usually known as tight energy-matter coupling, here applies to all heat fluxes in the system.

\begin{figure}[t]
\begin{center}
\includegraphics[width=0.7\linewidth,clip] {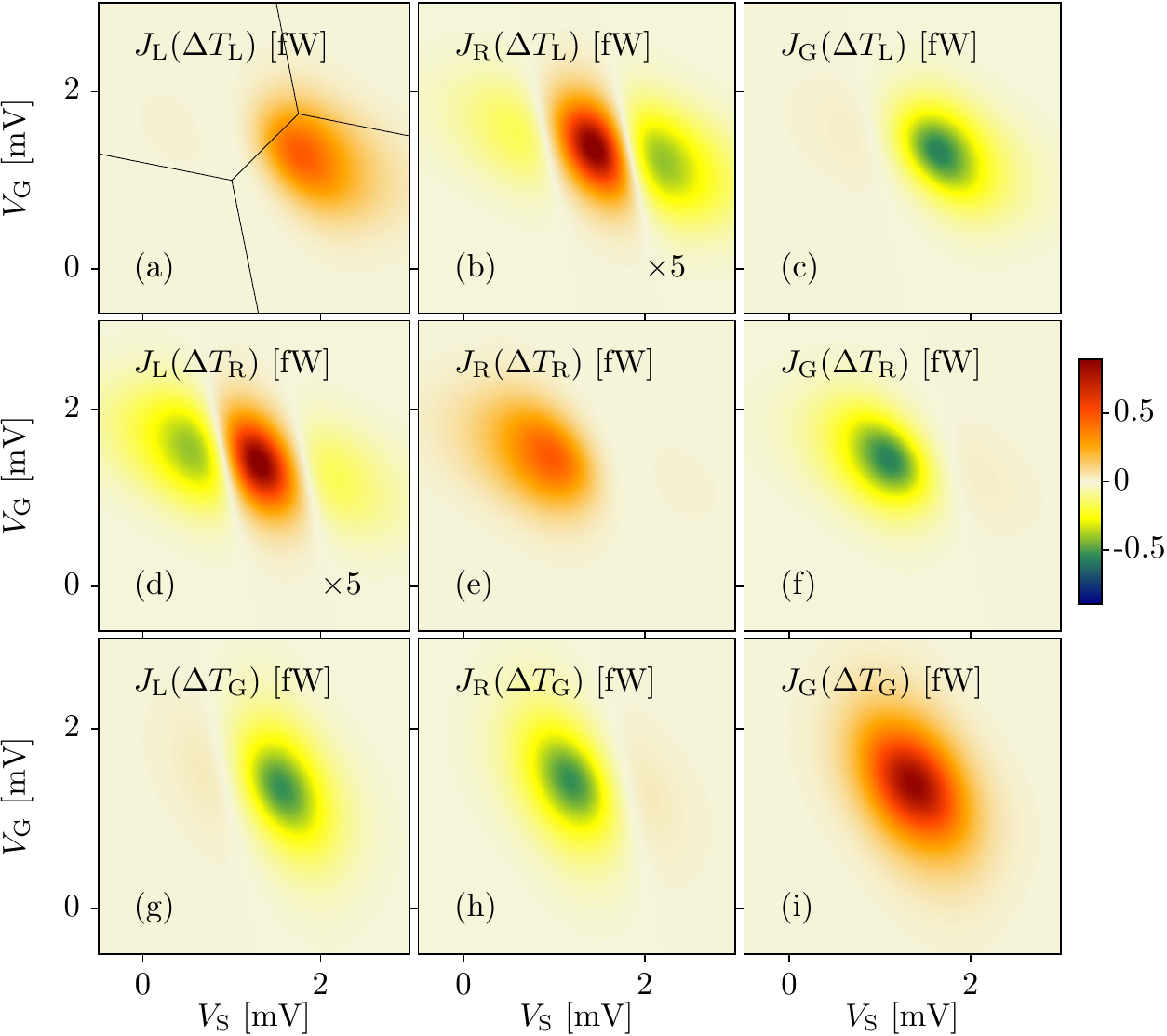}
\end{center}
\caption{\label{currentsOpt}Heat currents in response to the different temperature gradients as functions of the quantum dot gate voltages in the maximally correlated configuration, with $\Gamma_{\rmL1}=\Gamma_{\rmR0}=0$. Note that the heat currents $J_\rmL$, $J_\rmR$ change sign when the gradient is applieed to the opposite terminal in the conductor. In that case also, they are centered at different triple points. The rest of parameters is as in Fig.~\ref{currentsSym}.
}
\end{figure}

In Fig.~\ref{currentsOpt}, the different heat currents are displayed for the temperature gradient applied to the different terminals. They occur only around the stability vertex where energy can be exchanged with the gate. Each current is closer to a different triple point, corresponding to the  charge configuration at which tunneling is permitted. In agreement with Eqs.~(\ref{currsdiode}) and (\ref{partdiode}), when the temperature gradient is applied to one of the conductor terminals, $l$, the corresponding heat current $J_l$ vanishes (but does not change sign) when $\Delta U_{{\rm s}0}=0$ ($l$=L) or when $\Delta U_{{\rm s}1}=0$ ($l$=R). 

\subsection{Cooling by (really) heating}
Very remarkably, the current at the opposite terminal vanishes at two points, when $\Delta U_{{\rm s}n}=0$, where it changes sign. As a consequence, we find configurations in the center of the stability vertex where heat is extracted from the two terminals of the conductor, $J_\rmL,J_\rmR>0$, i.e. by heating one of them, the other one gets cooled. The excess heat is absorbed by the gate system. Hence only in the strictly hot terminal, our setup satisfies the Clausius statement that heat flows from hot to cold in the absence of work done onto the system. We also find heat flowing between two cold reservoirs.  Differently from other proposals of cooling by heating~\cite{mari_cooling_2012,cleuren_cooling_2012,chen_quantum_2012}, here we do not couple to an incoherent source that drives a particular transition, but rather really heat one part of the system up. 

Similar effects have been found for two finite systems coupled by energy filters~\cite{parrondo}. In our case, this is a dynamical effect due to the coupling to the gate, which enables the transported electrons to tunnel at different energies depending on the involved lead. Related properties have been described for asymmetric quantum dots in the dynamical Coulomb blockade regime~\cite{guillem}, suggesting the possibility to define thermal devices by properly engineering the coupling to their environment.

\begin{figure}[t]
\begin{center}
\includegraphics[width=0.7\linewidth,clip] {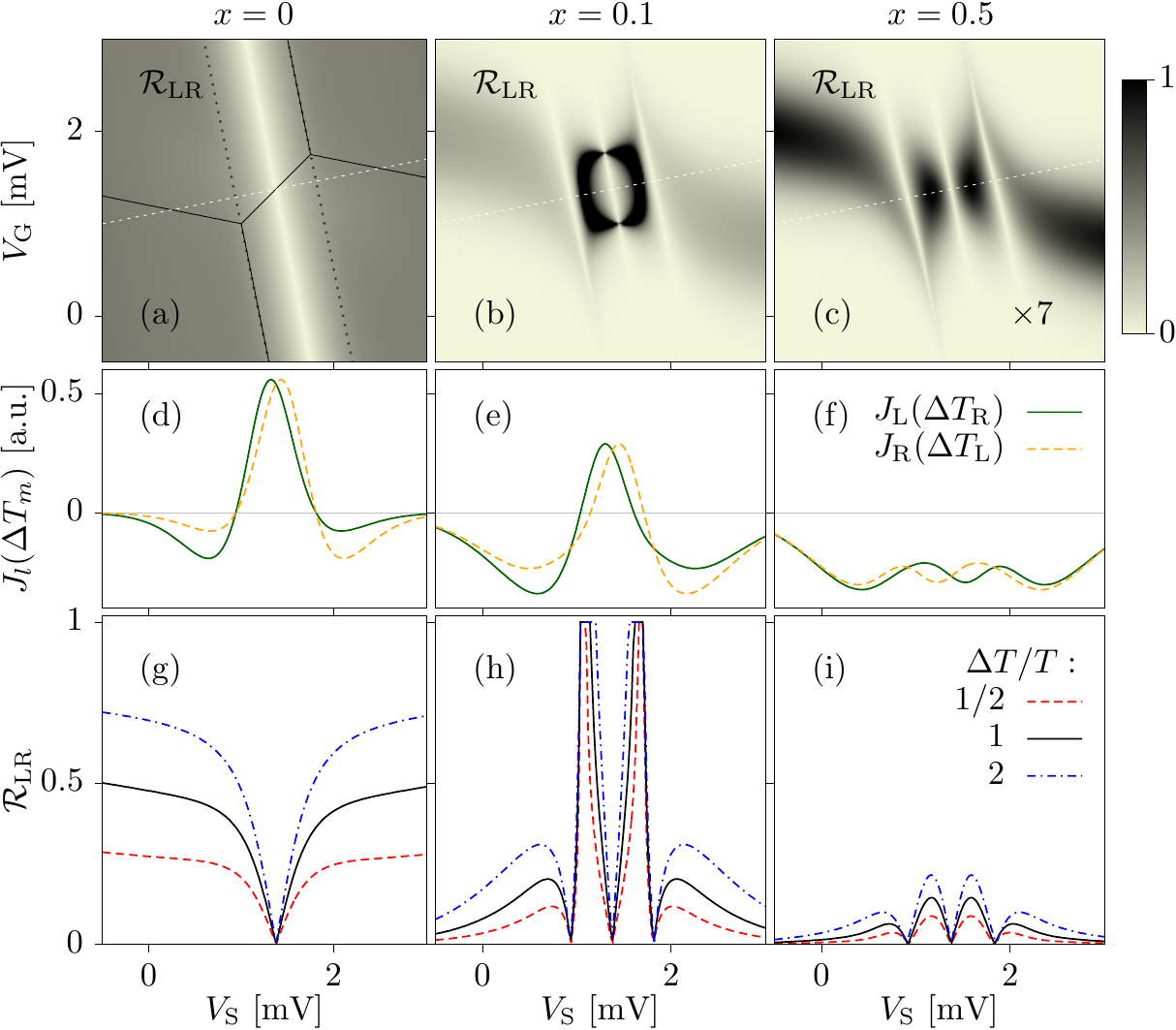}
\end{center}
\caption{\label{rectifcoeff}Thermal rectification, for the case with reversed energy-dependent tunneling rates: with $\Gamma_{\rmL1}=\Gamma_{\rmR0}=x\Gamma$. (a-c): Thermal rectification coefficient as function of the quantum dot gate voltages, for different values of $x$. The for the lower panels present cuts along the dashed white line: (d-e) represent the involved heat currents and (f-i) the dependence of the rectification coefficients for different values of the temperature gradient. Regions where ${\cal R}_{\rm LR}=1$ appear when the heat currents have opposite sign. Rest of parameters are otherwise as in Fig.~\ref{currentsOpt}.
}
\end{figure}

\subsection{Thermal rectification}
Let us concentrate on the thermal rectification. In Fig.~\ref{currentsOpt}, one can clearly observe an asymmetry of the heat currents due to an opposite gradient: $J_l(\Delta T_{\bar l})\neq J_{\bar l}(\Delta T_{l})$. It is due to the heat currents being largest at different dot potentials for different terminals. The thermal rectification coefficient will therefore be finite, except for the case when the system is equidistant from the conditions $\Delta U_{{\rm s}0}=0$ and $\Delta U_{{\rm s}1}=0$, as shown in Fig.~\ref{rectifcoeff} for $x=0$. It increases with the detuning of the conductor level with respect to the Fermi energy, where on the other hand the currents are exponentially suppresed. Interestingly, ${\cal R}$ increases with the applied temperature gradient, $\Delta T$, and gets closer to ${\cal R}=1$ for larger gradients.

\subsection{Leakage currents: thermal diode}
\label{sec:diodeleakage}

Remarkable deviations from this behaviour appear for $x\neq0$. In this case, a small current will leak at {\it undesired} energies at each of the terminals. This modifies the conditions at which the heat currents vanish. In particular, $J_\rmL(\Delta T_\rmR)$ and $J_\rmR(\Delta T_\rmL)$ change sign at different points, cf. Fig.~\ref{rectifcoeff} for $x=0.1$. As a consequence, two small voltage windows open when they are opposite in sign, i.e. they have the same direction. Within these windows, the system behaves as an ideal thermal diode, with ${\cal R}=1$. The size of the windows increases with the applied temperature gradient.

For larger leakage currents, the cooling by heating effect disappears, so both $J_\rmL(\Delta T_\rmR)$ and $J_\rmR(\Delta T_\rmL)$ are negative, cf. Fig.~\ref{rectifcoeff} for $x=0.5$. They are still different, in general, and hence a finite but small rectificaction coefficient is obtained.

%\section{Thermal cooling}
\section{Quantum dot arrays for energy filtering}
\label{sec:filters}

We end by analyzing the experimental feasibility of the different configurations discussed above. The energy-dependent rates of systems of two coupled quantum dots can be modified by means of gate voltages~\cite{holger}. However, even if configurations with large enough asymmetries can be found~\cite{bischoff_measurement_2015}, they are very difficult to control.

\begin{figure}[t]
\begin{center}
\includegraphics[width=0.7\linewidth,clip] {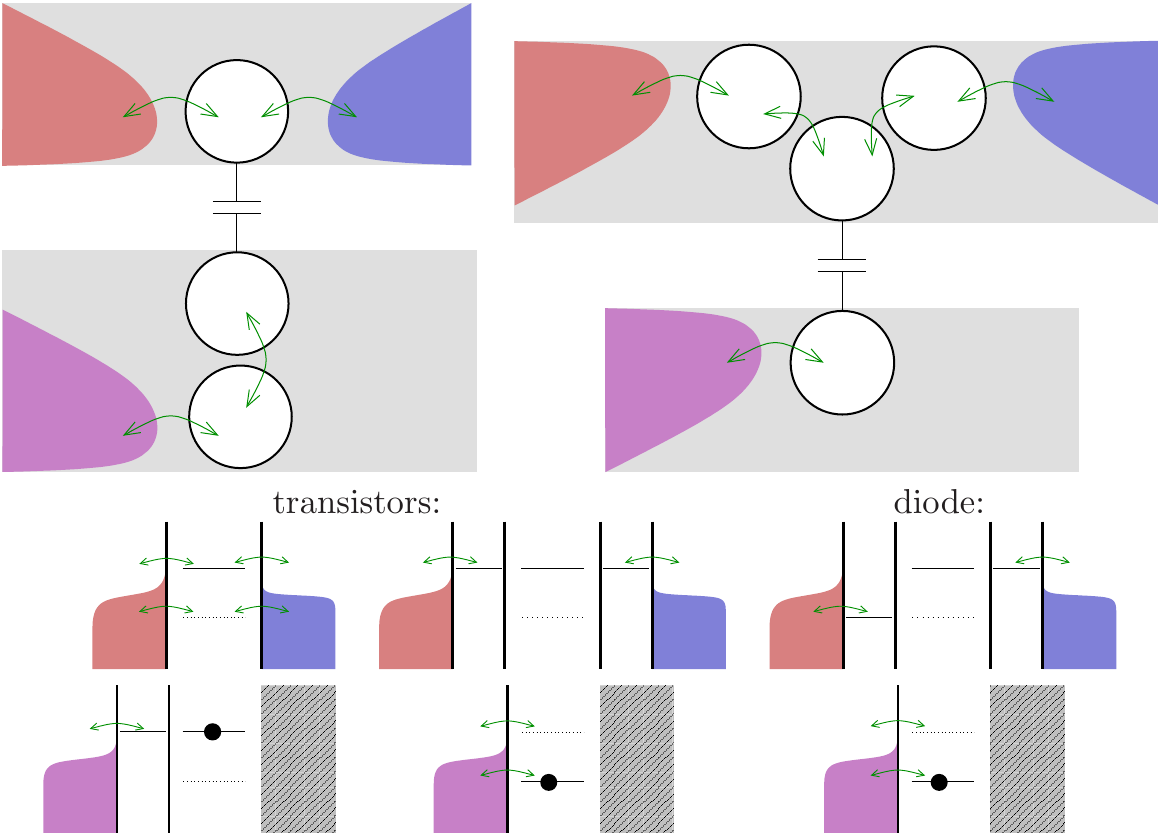}
\end{center}
\caption{\label{filters}Possible realizations of energy selective tunneling with quantum dot arrays. The coupling between the system and the gate is given by the capacitive coupling of only two dots. The additional dots filter the energy at  which they are coupled to the reservoirs. Different operations as a thermal transistor or a thermal diode can be defined this way.
}
\end{figure}

One can however think of more complicated quantum dot arrays. The discrete level of the outermost quantum dots can be used as energy filters which can be tuned externally. The coupling between the conductor and gate subsystems is mediated by only two dots, as sketched in Fig.~\ref{filters}. This way, tunneling from the different terminals can be put in resonance with the required configurations at $\Delta U_{im}$, as long as the width of the filter levels is smaller than $E_C$~\cite{rf}.

The simplest case, as it only involves one additional barrier, is when the gate contains a double quantum dot. The one connected to the reservoir behaves as a zero-dimensional contact~\cite{bryllert_designed_2002}. Its level can be chosen to inject electrons only either at $\Delta U_{\rmG0}$ or $\Delta U_{\rmG1}$, thus providing a configuration with either $\Gamma_{\rmG1}=0$, or with $\Gamma_{\rmG0}=0$, respectively. They would work as thermal transistors around different triple points in the stability diagram~\cite{short}.

Controlling tunneling in the conductor (as discussed in this work) would require a triple quantum dot, cf. Fig.~\ref{filters}. Fine tuning of such structures has been achieved in the last decade~\cite{gaudreau_stability_2006,schroer_electrostatically_2007,grove-rassmussen_triple_2008,rogge_quantum_2013}. On the other hand, it permits for a more flexible operation of the system: if the two filters are resonant with the same state, $\Delta U_{ln}$, the rates with $n'\neq n$ will vanish ($\Gamma_{ln'}=0$), and the system is a thermal transistor, as discussed in Sec.~\ref{sec:transistor}. If now one of the filters is put in resonance with the other state, it works as a thermal diode or a refrigerator, see Sec.~\ref{sec:diode}. We can therefore change the operation of the device from a transistor to a diode by just changing a gate voltage.

\section{Conclusions}
\label{sec:conclusion}

We have investigated the properties of electronic heat transport in a system of capacitively coupled quantum dots in a three-terminal configuration. One of the dots acts as a mesoscopic thermal gate which can either serve a a heat source or sink, or control the occupation of the other dot. Heat exchange mediated by electron-electron interactions introduces different ways to tune the energy-dependence of the relevant tunneling events. 

All-thermal control of heat currents can be achieved by controlling the appropriate system asymmetries. It allows us to define different operations such as a thermal transistor, a thermal diode, and a refrigerator based on the interaction with the gate system. On one hand, dynamical channel blockade of the conductor heat currents via charge fluctuations in the gate enable a thermal transistor with huge amplification factors. On the other hand, the correlation of state-dependent transitions in the three different contacts rectifies the heat currents and induces the cooling of one terminal by the heating of the other one. These two effects combine in the operation of an ideal thermal diode with ${\cal R}=1$. 

The different configurations can be implemented under nowadays experimental state of the art. In arrays of quantum dots, the optimum level of performance can be achieved. Our results consider realistic parameter estimations from the experiment~\cite{holger}.

Our work uses a mesoscopic structure to mediate the coupling of the conductor with the thermal environment. Working as a switch or inducing inelastic transitions, the mediator defines the different behaviours of the system. This strategy opens the possibility to use engineered system-environment couplings to improve or define new functionalities.

\ack 
We acknowledge financial support from the Spanish Ministerio de Econom\'ia y Competitividad via grants No. MAT2014-58241-P and No. FIS2015-74472-JIN (AEI/FEDER/UE), and the European Research Council Advanced Grant No. 339306 (METIQUM). We also thank the COST Action MP1209 ``Thermodynamics in the quantum regime". 
%\end{acknowledgments} 

\end{document}